# I don't trust you (anymore)! – The Effect of Students' LLM Use on Lecturer-Student Trust in Higher Education

SIMON KLOKER[1,2,*] (0000-0003-3653-9836),
MATTHEW BAZANYA[1] (0009-0001-9612-4387)
AND TWAHA KATEETE[1] (0009-0006-9532-7154)
[1]*Ndejje University, Faculty of Science and Computing, Uganda*
[2]*Coworkers International, Germany*
**Corresponding author: skloker@ndejjeuniversity.ac.ug*

Accepted 9 September 2024, published 25 September 2024

*https://doi.org/10.58653/nche.v12i1.6*

## Abstract

Trust plays a pivotal role in lecturer-student collaboration, encompassing teaching and research aspects. The advent of Large Language Models (LLMs) in platforms like Open AI's ChatGPT, coupled with their cost-effectiveness and high-quality results, has led to their rapid adoption among university students. However, discerning genuine student input from LLM-generated output poses a challenge for lecturers. This dilemma jeopardizes the trust relationship between lecturers and students, potentially impacting university downstream activities, particularly collaborative research initiatives. Despite attempts to establish guidelines for student LLM use, a clear framework that is mutually beneficial for lecturers and students in higher education remains elusive. This study addresses the research question: How does the use of LLMs by students impact Informational and Procedural Justice, influencing Team Trust and Expected Team Performance? Methodically, we applied a quantitative construct-based survey, evaluated using techniques of Structural Equation Modelling (PLS-SEM) to examine potential relationships among these constructs. Our findings based on 23 valid respondents from Ndejje University indicate that lecturers are less concerned about the fairness of LLM use per se but are more focused on the transparency of student utilization, which significantly influences Team Trust positively. This research contributes to the global discourse on integrating and regulating LLMs and subsequent models in education. We propose that guidelines should support LLM use while enforcing transparency in lecturer-student collaboration to foster Team Trust and Performance. The study contributes valuable insights for shaping policies enabling ethical and transparent LLMs usage in education to ensure the effectiveness of collaborative learning environments.

**Keywords:** *Large Language Models; lecturer-student collaboration; Informational Justice; Team Trust; Structural Equation Modelling.*

# Introduction

"Hello GPT! Goodbye home examination?" is the title of a recent publication that succinctly summarizes the dilemma faced by millions of teachers and lecturers worldwide (Farazouli et al., 2024). Trust in students' unique work erodes rapidly when there is this pervasive concern that it might be simply generated using artificial intelligence (AI), more precisely a Large Language Model (LLM). In the dynamic landscape of higher education, the relationship between lecturers and students plays a pivotal role in fostering collaborative learning environments and research. Resilient trust relationships between lecturer and student empowers both parties: The lecturer gains a motivated and educated assistant in his research undertakings, while the interaction and joint work not only motivate students to actively participate but also cultivates essential interpersonal and professional skills (Laal & Laal, 2012; Mendo-Lázaro et al., 2022; Yang, 2023). Trust serves as the bedrock for the success of this collaborative research endeavour (Lewicka & Bollampally, 2022).

However, recent trends of increasing use of LLMs among students in higher education have jeopardized this relationship. LLMs, exemplified by ChatGPT, Google Bard (Gemini), or Meta's LLaMA, have become integral tools for students, aiding in various tasks such as assignment development, essay writing, and coding (Gamage et al., 2023; Ward, 2023). Trained on tons of data, LLMs can write, read, do research and respond like humans (Dwivedi et al., 2023). While the benefits of LLMs are acknowledged, concerns regarding academic integrity, quality issues, and the potential misuse of generated content have arisen (Lo, 2023; Strzelecki, 2023; Sullivan et al., 2023). The ethical use of LLMs becomes crucial to avoid compromising the integrity of educational practices. In addition, they put the lecturer in a weaker position, where he no longer can be sure that – whatever students deliver to them – is their work and created with academic integrity.

While most of the current discussion on the entry of LLMs in higher education focuses on its ethical implications, fairness, and its impact on educational issues such as a tool for new didactic concepts, most of these works focus on individual settings, such as homework, essays, or individual learning. Few, if any, mention how LLMs' presence affects the bidirectional working and trust relationship in teams of students and lecturers – although it is well documented that trust in this relationship is a crucial element for performance in teaching and research.

This paper explores the impact of students' LLM use on the lecturer-student trust relationship in higher education. The rise of LLM use introduces a complex interplay between technological advancements, ethical considerations, and the traditional foundations of trust in education and research. Understanding this dynamic is essential for shaping guidelines and strategies that ensure the positive integration of LLMs in the learning and research process.

# Related Work and Research Gap

## Trust in Lecturer-Student Collaboration

Lecturer-student collaboration, also referred to as collaborative learning (Laal & Laal, 2012; Yang, 2023), is defined by Laal and Laal (2012) as an umbrella term encompassing various educational approaches involving joint intellectual efforts by students, either independently or in conjunction with teachers. It serves as a crucial mechanism in academic research, as teams consisting of lecturers and students are common in the context of higher education, although lecturers will mainly play a guidance role. Research project-based teaching approaches that motivate students to actively engage in the learning and research process also foster the development of interpersonal skills (Mendo-Lázaro et al., 2022). Seminar papers and theses represent two manifestations of such teaching approaches, empowering students to decide what, when, and how to study, under the guidance of the lecturer, thereby collectively shaping the learning experience and outcomes. Abbas et al. (2020) argue that there is a direct relationship between teacher-student trust dynamics and collaborative research, as the latter relies heavily on the former.

For any relationship to thrive, trust must be paramount (Lewicka & Bollampally, 2022). Hence, in the context of lecturer-student collaboration, positive outcomes necessitate both parties being trustworthy and trusted. Bain (2004) found, based on research on college teachers, "[…] that 'highly effective teachers' cultivated relationships characterised by openness with and trust in students" (Woodland & Woodland, 2011, p. 5). Similar sentiments are echoed by Florén (2003). Carless (2009) emphasizes the negative impact of distrust on performance, especially in learning-oriented assessment practices such as seminar papers or theses.

Beyond the educational context, there is additional evidence highlighting the interconnectedness of Trust, Collaboration, and Team Performance. Teams, in general, encompass various roles (Schmutz et al., 2019), both in academia and industry (where most research on teams originate). In the academic context, the teacher who acts as a facilitator/guide works closely with a team of students or a team leader, typically chosen by fellow students. Similarly, in regular work teams, the unit manager collaborates closely with a team leader to ensure the accomplishment of assigned tasks. In both scenarios, trust forms the foundation, as teams cannot deliver the expected outcomes without it (Ji & Yan, 2020).

## Rise of LLM-Use among Students in Higher Education

The usage of LLMs among students in higher education is continually increasing. LLMs have undergone a revolutionary advancement in conversational AI and swiftly established positions in academia (Strzelecki, 2023). In consideration of their distinctive user-friendly features, LLMs like ChatGPT have garnered much attention and sparked controversy among students and academics ever since their release (Gamage et al., 2023). Many scholars foresee them becoming as ubiquitous as or even more ubiquitous than other famous search engines (Hack, 2023). Students leverage LLMs for various tasks, including producing texts, developing assignments, supporting essay writing, providing responses to questions, coding, and other tasks (Gamage et al., 2023). Ward's (2023) survey of 1,000 U.S. college students revealed that nearly 90% used ChatGPT for assignments or essays, aligning with the experiences reported by many lecturers nowadays.

On the positive side, literature highlights several beneficial effects of LLMs in higher education. These include increased engagement in Massive Open Online Courses (MOOCs) (Li & Xing, 2021), enhanced remote learning and collaboration (Lewis, 2022), and the creation of interactive, gamified, more tailored, or personalized assessment and teaching methods (Cotton et al., 2023; Ilieva et al., 2023; Kloker et al., 2024; Sullivan et al., 2023). Conversely, a significant portion of the discussion revolves around concerns regarding quality and academic integrity (Gamage et al., 2023; Strzelecki, 2023; Sullivan et al., 2023). Thereby, most authors agree that it is fair for students to use LLM-generated responses as a starting point for their solutions or as a guide to building up well-structured, grammatically correct, and complete answers – as long as they augment them with their own ideas, knowledge, and critically validate the content (Lo, 2023). Failure to incorporate this "human loop" can lead to various problems, which Lo (2023) summarized, among others, in these three categories based on a short literature review: (1) Quality issues, where LLMs may deliver results based on biased data or invent information and citations; (2) Integrity issues, where students may claim LLM-generated content as their own or use it to circumvent plagiarism detection by using LLMs to rephrase original pieces of work (e.g., Cotton et al., 2023); and (3) Educational issues, where testing and assessing (as well as developing) of higher-order critical thinking of students may require new formats (e.g., Hsiao et al., 2023).

The second issue, in particular, raises concerns among academics (Gamage et al., 2023; Lo, 2023; Sullivan et al., 2023). The willingness and readiness of students to adopt these technologies is evident from lecturers' experiences and literature (Abdaljaleel et al., 2024; Cotton et al., 2023; Hsiao et al., 2023; Ward, 2023). However, prohibiting LLMs using technical means is deemed impractical (Sullivan et al., 2023), as is detection in the long run (Gorichanaz, 2023). The challenge lies in the ambiguous distinction between fair and unfair LLM use (Perkins, 2023; Roe & Perkins, 2022) – which is true for both lecturers and students. However, they put lecturers in a weaker position as they might no longer be able to assess students' work properly, bearing the question in mind whether hand-ins are authentically students' solutions. Consequently, this negatively affects the trust relationship in lecturer-student collaboration.

## Derivation of Research Question

The use of LLMs in the context of higher education is still confronted with ambiguous and contradicting views (Gamage et al., 2023). Worldwide, universities grapple with establishing consistent policies and guidelines, ranging from embracing to outright banning of this technology. Students, too, perceive this ambiguity and fear that their use of LLMs might adversely affect the lecturer-student collaboration and relationship (Ofosu-Ampong et al., 2023; Singh et al., 2023). The rise of LLM use by students and its inherent ambiguity create a research gap, whether this new collaboration setting also affects team trust. While previous research primarily focused on the aspect of fairness (or cheating, respectively), the current study aims to investigate the overall impact, both positive and negative, on expected performance beyond "single-authored" essays.

In our study, we employ *Team Trust* as a central construct in lecturer-student collaboration, given its well-documented positive impact on *Team Performance*, as evidenced by a meta-analysis of 112 studies conducted by De Jong et al. (2016). It is plausible to suggest that *(Perceived) LLM Usage* by students may erode this *Team Trust*. In 2023, several studies that suggest this relationship have been published. For instance, Joshi et al. (2023) conducted interviews among lecturers (instructors) and students, revealing that especially instructors are worried regarding their ability to assess students' work any longer, as they cannot distinguish between LLMs output and actual student work.

Similar concerns are echoed, among others, by Perkins (2023). Concurrently, lecturers are aware that students perceive low risk in terms of detection for academic misconduct or plagiarism (Abdaljaleel et al., 2024; Gorichanaz, 2023). This causes anxiety among the lecturers and leads to accusations, whether valid or not (Gorichanaz, 2023) – thereby straining the trust relationship in lecturer-student collaboration (Grassini, 2023). Grassini (2023) already suggests that these concerns and a reduced ability to assess students' work fairly may undermine the effectiveness of the educational process.

Aligned with Costa and Anderson (2011), we argue that *Team Trust* is a multifaceted formative construct[1], implying that the relationship between *(Perceived) LLM Usage* might affect the individual reflective sub-constructs of *Team Trust* differently. Building on the work of Breuer et al. (2020), we identify perceived justice as an antecedent for *Team Trust*. We suppose that there might be a mediating effect of *Informational Justice* on *Perceived Trustworthiness*, as well as on *Cooperative Behaviour* (both reflective sub-constructs of *Team Trust*), whereby their effect on trust in the context of teams is already demonstrated by Kernan and Hanges (2002) and Palanski et al. (2011) (although on *Trust in Management* and overall *Team Trust*, respectively). We argue that this is an important mediation, reflecting the frequently reported information asymmetry regarding the origin of students' work outcomes when LLM use is assumed.

As another mediating path, we argue that *Procedural Justice* impacts both *Cooperative Behaviour* and *Monitoring Behaviour*, reflecting the existing ambiguity surrounding whether students' use of LLMs is a valid method or a violation of academic practices. In the absence of clear guidelines on fair LLM Usage in many universities, lecturers become the arbiters of this question. The effect of *Procedural Justice* on *Team Trust* is suggested by Colquitt (2004) and Dayan and Di Benedetto (2010). We also propose a mediating effect, though we recognize a valid justification for considering it as a moderating effect on the relationship between *(Perceived) LLM Usage* and *Team Trust*.

Exploring these relationships is of great importance in providing guidelines for future lecturer-student collaboration in environments where LLMs are not banned. We put the following explorative research questions: *How does the use of LLMs by students impact Team Trust and subsequently Expected Team Performance? (RQ1) What is the role of Informational Justice and Procedural Justice in this relationship? (RQ2)* We aim to shed light on these relationships and propose a research model and potential extensions for statistical and structural validation in a future study.

# Methodology and Implementation

To identify potentially relevant relationships among the proposed constructs in Section 2.3, we conducted an online questionnaire. The survey comprised four sections: (1) An introduction that briefly explained the importance of the research to participants and requested thoughtful and truthful commitment. (2) The second part inquired about the main constructs in three blocks, with 10 questions per block and intra-block randomization of items. Two manipulation checks were included in these blocks. (3) The third part gathered demographics and additional control variables. (4) The final section concluded

1 Reflective sub-constructs are: Prospensity to Trust, Perceived Trustworthiness, Collaborative Behaviour, Monitoring Behaviour (Costa & Anderson, 2011) the survey with a short debriefing. Table 2 in the Appendix displays the final items in the online questionnaire.

The questionnaire items were measured using a 5-point Likert scale from "Strongly disagree" to "Strongly agree". This was the original scale for the constructs of *Informational Justice* and *Procedural Justice* (Colquitt, 2004; Dayan & Di Benedetto, 2010; Kernan & Hanges, 2002; Palanski et al., 2011). The formative construct of *Team Trust* from Costa and Anderson (2011), originally based on a 7-point Likert scale with similar labels, was easily adapted to the 5-point scale. Employing the same scale for all constructs allowed us to present all items within three clear blocks with intra-block randomization and two manipulation checks. The overall survey comprised 38 questions, an introduction, and a short debriefing.

The survey was implemented using the questionpro.com software and distributed among approximately 200 university lecturers at Ndejje University (Uganda) via WhatsApp between January 10th and January 14th, 2024. Participation was voluntary, and no incentives were offered, although we emphasized the supportive and positive impact of participants' answers on our research during the invitation.

A total of 32 responses were collected (see Section 5 for a discussion of the low return rate). The survey included two manipulation checks and we only considered responses that successfully passed the second check[2], resulting in 23 remaining respondents, including nine females. The average completion time among the included responses was 13 minutes and 55 seconds. The median age of participants fell within the range of 35 – 44 years, with more than 95% of participants aged between 25 and 54 years. Eighty-seven per cent had already supervised students; however, half of them had less than 41 students. This also reflects the fact that almost 70% of respondents would assess their research experience as rather low or medium. Seventy-four per cent of respondents reported having used LLMs at least once. Only 30% reported having had negative experiences with students' LLM use so far, and almost all of them more than once. Forty per cent of respondents teach in ICT, 13% in Social Sciences, with others scattered among various fields.

# Results

As part of the analysis of the exploratory research objective, we employed PLS-SEM for data analysis. We only retained constructs with Cronbach's alpha (Cron. $\alpha$) values that exceeded the threshold of 0.6. After tuning Cron. $\alpha$ by excluding items IJ1, PJ1, PW3, and TP1, we omitted *Propensity to Trust* (Cron. $\alpha$ = .56), *Collaborative Behaviour* (Cron. $\alpha$ = .24), and *Monitoring Behaviour* (Cron. $\alpha$ = .21).

For the remaining analysis, we followed the two-stage approach outlined by Hair et al. (2021). First, we assessed internal consistency reliability, convergent validity, and discriminant validity. For all retained constructs, both Cronbach's $\alpha$ and composite reliability (CR) exceeded the threshold values of 0.6 and 0.7, respectively. Subsequently, we evaluated convergent validity by examining each item's outer loading and the average variance extracted (AVE) for each construct. Most of the outer loadings exceed the threshold of 0.7. For the others, we still opted not to remove them, as, according to Hair et

2 Respondents feedbacked, that the first manipulation check seemed to confuse participants - and was therefore excluded.

al. (2021), items with an outer loading between 0.4 and 0.7 should only be eliminated when AVE or measures of internal consistency and reliability move above the cut-off value, which we verified to not be the case. AVE for all constructs was above the cut-off value of 0.5, except for *Informational Justice* (IJ, .497) and *Perceived Trustworthiness* (PW, .472), which only slightly fell below the threshold. To assess discriminant validity, we employed three checks: consideration of cross-loadings, the Fornell-Larcker criterion, and the Heterotrait-Monotrait Ratio (HTMT). All three approaches provided supporting evidence of the discriminant validity of the constructs.

Table 1 presents the construct reliability for the reflective constructs in the survey, after adjusting Cron. $\alpha$ by dropping items IJ1, PJ1, PW3, and TP1.

***Table 1:*** Reliability Measures of the Included Constructs (Cronbach's Alpha)

| Construct | Mean | SD | Cron. $\alpha$ | CR | AVE | HTMT* |
|---|---|---|---|---|---|---|
| (Perceived) LLM Usage (LU) | 3.03 | 1.22 | 0.83 | 0.97 | 0.55 | No |
| Informational Justice (IJ) | 3.20 | 0.95 | 0.75 | 0.80 | 0.50 | No |
| Procedural Justice (PJ) | 2.63 | 1.07 | 0.62 | 0.82 | 0.54 | No |
| Perceived Trustworthiness (PW) | 3.32 | 0.93 | 0.68 | 0.82 | 0.47 | No |
| Team Performance (TP) | 3.07 | 1.04 | 0.72 | 0.82 | 0.61 | No |

*Indicates whether HTMT CI includes 1.

Secondly, we checked for collinearity issues among predicting constructs and found that all VIF values are well below the threshold of 5, indicating the absence of collinearity issues[3] in our structural model.

To explore potential relationships in the data, we correlated the remaining constructs in Figure 1:

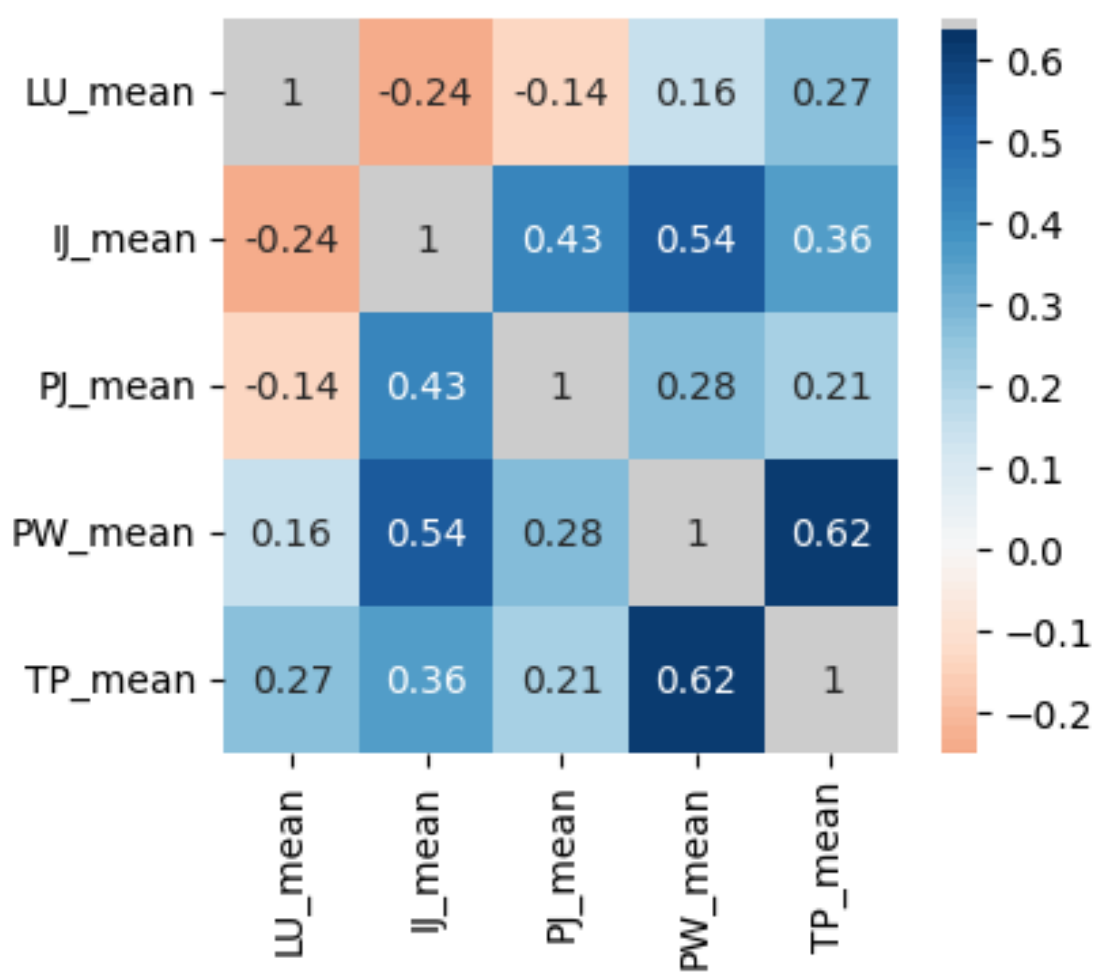

***Figure 1:*** Correlation Matrix

*(Perceived) LLM Usage* (LU) is weakly negatively correlated with IJ and *Procedural Justice* (PJ), aligning our expectations based on literature. IJ and PJ exhibit positive correlations with PT and *Team Performance (TP)*; also, in line with our hypotheses. LU is weakly positively correlated with PT, suggesting that the effect may be either mediated or moderated by IJ and PJ. Interestingly, LU is positively correlated with TP implying that, although the lecturer's perception of Team Trust is reduced, it does not negatively affect their expected team results.

In addition, the control variable "How often do you use Large Language Models, such as ChatGPT?" (C3) has a big influence (+.55, R2=0.31) on LU. Previous negative experiences with students' LLM use (C5) also demonstrate strong predictive power on LU (-.52, R2=.28 – note that it was not measured on a metric scale). The control variable assessing whether lecturers perceive LLMs as a fair tool in general (C8) has a substantial effect on IJ (note that it was measured as a binary). Consequently, we model it as a potential moderator for LU→IJ and LUPJ. The resulting SEM is illustrated in Figure 2.

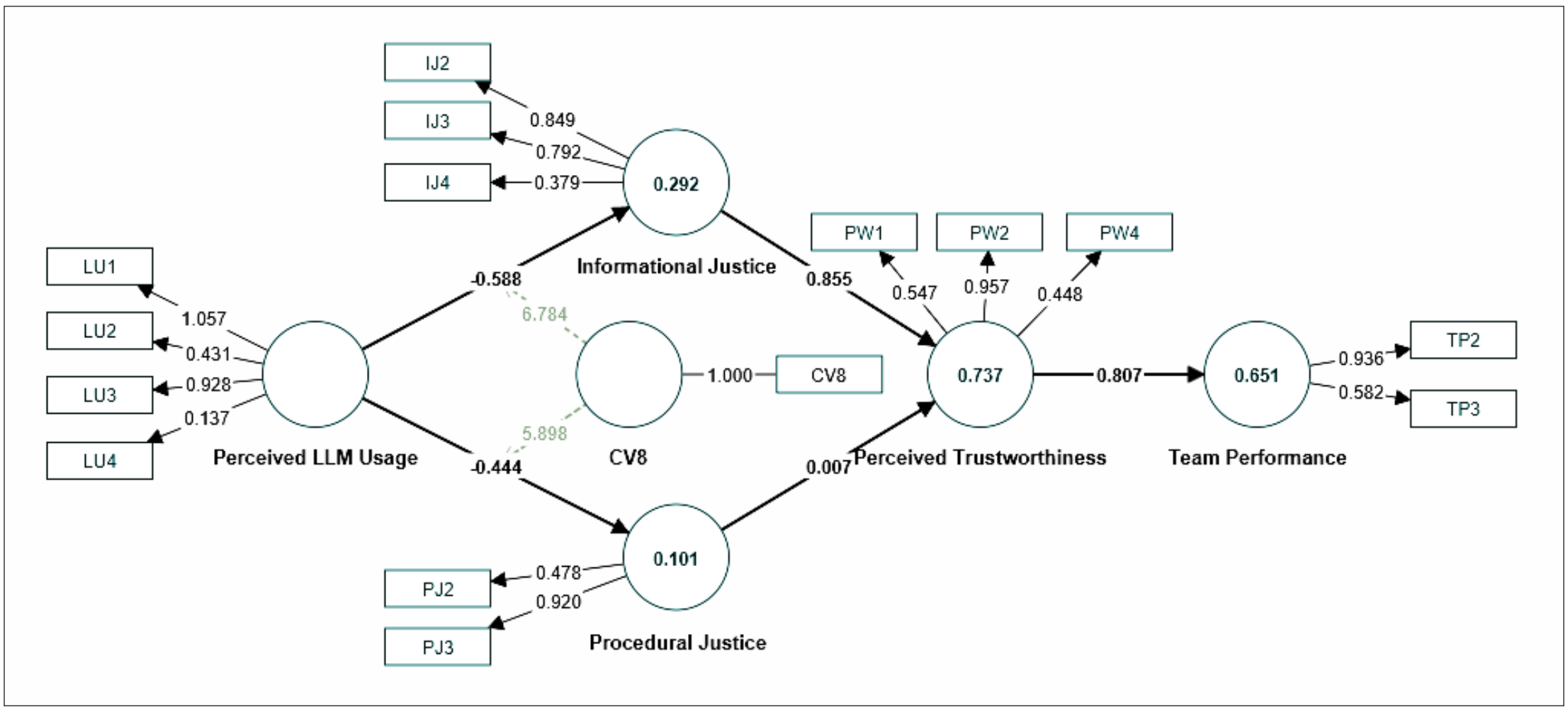

***Figure 2:*** Suggested Structural Equation Model

Owing to the limited sample size, we refrain from conducting a significance analysis using bootstrapping. We also try to avoid overinterpreting the model, and merely highlight the most interesting tendencies for follow-up verification.

Firstly, when the moderating variable C8 is omitted, both path coefficients LU→IJ and LU→PJ are very low (below 0.1), as well as the corresponding $R^2$ values (below 0.1). Only after adding the moderator variable, path coefficients and $R^2$ values for IJ and PJ increase to relevant levels, although PJ's $R^2$ still remains low. Secondly, the path coefficient between IJ and PW is strong and already significant despite the small sample size, while the path coefficient between PJ and PW remains almost zero. Thirdly, the relationship between PW and TP aligns with expectations from the literature.

# Discussion

As a first finding, LU tends to have a positive effect on *Expected Team Performance* (TP), suggesting – despite all the discussion regarding appropriate and fair use – positive expectations regarding students who use LLMs being able to produce better overall results (RQ1). This finding is in line with the arguments presented in several papers (e.g.,

Gamage et al., 2023; Gimpel et al., 2023; Roe & Perkins, 2022). However, there seems to be no quantitative evidence yet of this increased Team Performance for lecturer-student collaboration employing LLMs, such as seminar papers or theses, beyond "self-studies". Although our study suggests the same effect, it is crucial for future studies to evaluate this performance increase not in isolation but in conjunction with its (positive or negative) effects on expected teaching success (Gamage et al., 2023). To answer RQ1, it can be stated that the effect of LU on TP is positive.

As a second finding, based on the evaluation, the data suggests that *Perceived LLM Usage* (LU) by students is actually accepted by lecturers. This acceptance is evident not only in the control variable C8, where 87% of respondents answered that they perceive LLMs as fair tools in the context of higher education, but also in the Structural Equation Model (SEM), where the effect of *Procedural Justice* (PJ) to *Perceived Trustworthiness* (PW) is almost zero, and only the path LU→*Informational Justice* (IJ)→PW shows relevant effects. This suggests that the more crucial issue is that students make their use of LLMs transparent to the lecturer, particularly in cases where lecturers have doubts regarding the fairness of using LLMs. The high path coefficients argue towards the interpretation that this transparency moderates the relationship between LU and IJ. This finding aligns with suggestions by Gimpel et al. (2023) or Lo (2023), providing a great deal of evidence where lecturers acknowledge the positive usage scenarios of ChatGPT as long as transparency is ensured. To answer RQ2, it can be stated that PJ seems not to be relevant, while IJ influences the TP positively; however, LU has currently influenced IJ negatively. In other words, lecturers assume that students use LLMs secretly, although they would not have a problem with it if it was used transparently. They even expect better Team Performance if students would be open about their use.

As a third finding, we note that although the effect between PW and PT is as expected, it is essential to highlight that the applied *Team Trust* measure by Costa and Anderson (2011) performed poorly in our context. This might be due to several reasons, which need consideration in a follow-up assessment. *Team Trust* in the context of Costa and Anderson (2011) is a bidirectional construct, meaning that several items ask for two-sided trust relationships (e.g., PW1, PW2, CB1, CB3). Our study assessed only a unidirectional trust relationship from lecturers towards students. Future studies should reconsider the measure for trust or adapt the framing to yield better results. Overall, many of the measures for validity and reliability suggest that several items need to be revised, as some of them showed ceiling effects, such as those from *Monitoring Behaviour* and *Collaborative Behaviour*, which is likely the reason for their poor performance in reliability measures (Gaudet et al., 2021). It is also worth noting that although the effect between LU→IJ and LU→PJ was relevant, the $R^2$ for these constructs remained low, indicating that we missed out on relevant factors in these relationships. Future research should put emphasis here to further deconstruct the underlying principles and processes. RQ2 could, therefore, not be finally answered. However, we suggest a positive, by IJ-moderated relationship LU→PT.

One limitation of our study is that we omitted to include a theoretically unrelated construct to check for the Common Method Variance (CMV) (Podsakoff et al., 2003). As we did not offer incentives to our participants, keeping the survey short was regarded as the higher priority. Instead, we included the construct *Propensity to Trust*, which, as a personality trait, is not theoretically related to the hypotheses in the left part of the model and can, therefore, be used as a provisional marker variable to estimate the effect of CMV. However, the construct turned out to be unreliable in our survey, for which reason we could not test for CMV. In future studies, we suggest the inclusion of the measure of *Life Satisfaction* by Diener et al. (1985), as suggested by Simmering et al. (2015) for this reason.

Another limitation is that we did not reach the necessary sample size to test for effect significance. We conducted a power analysis according to Cohen (1988), anticipating a large effect size and accepting a probability level of 0.1 (power level of 0.8) upfront. The sample size needed to detect effects was 40, and 116 to confirm the model structure. Therefore, we aimed for 200 contacts, expecting at least a 30% response rate, ideally 50%. However, this assumption proved incorrect. We subsequently asked individuals thereafter about their reasons for participating or not. The most common negative responses were: (1) Impersonal approach ("Not willing to spend time or effort for somebody I don't know."); (2) low interdependence among lecturers ("Not willing to spend time or effort, as I don't expect to need similar assistance in future."); and unfamiliarity with surveys ("I didn't know what to do"). Future research should involve more universities to increase the reach and sample size. Also, incentives might be considered. Several self-identified non-respondents later indicated that an introduction of the survey within an online meeting, allowing for immediate question clarification, would have been helpful.

A third limitation is that, for now, we only considered the context of Uganda, while the context of the literature we based our hypotheses on was in other geographical areas. However, trust and overall attitude towards technology are strongly dependent on culture and individual settings (Abdaljaleel et al., 2024; Dinev et al., 2006; Kloker et al., 2020; Peukert & Kloker, 2020). For this reason, it might be necessary to contextualize the theory and findings within this new context and validate some of the assumptions within a Ugandan setting once again.

# Conclusion

In conclusion, this study delves into the intricate relationship between students' utilization of Large Language Models (LLMs) and the trust dynamics within lecturer-student collaboration in higher education. The escalating adoption of LLMs, like ChatGPT, introduces a paradigm shift, both creating opportunities and posing challenges for the educational landscape.

Our findings shed light on several critical aspects. Firstly, the acceptance of LLM Usage by lecturers suggests a nuanced perspective that recognizes the potential benefits as long as transparency is maintained. The importance of students making their use of LLMs transparent emerges as a pivotal factor moderating the relationship between LLM Usage and perceived Informational Justice. This nuanced perspective aligns with the acknowledgment of positive usage scenarios of LLMs, contingent upon transparent practices (Gimpel et al., 2023; Lo, 2023). This finding recommends that universities should rather encourage students to make their use of LLMs transparent instead of banning the technology or ignoring it.

Secondly, despite the ongoing discourse regarding the appropriate and fair use of LLMs, our study indicates a positive association between LLM Usage and expected Team

Performance. This suggests an optimistic outlook among lecturers, anticipating enhanced overall results when students employ LLMs. This finding resonates with previous arguments about the potential benefits of LLMs in educational settings (Gamage et al., 2023; Roe & Perkins, 2022). However, it calls for a nuanced evaluation of performance increases in the context of lecturer-student collaboration, especially in tasks such as seminar papers or theses, to be verified. This finding recommends that universities can even expect better overall results, when they allow and teach proper use of LLMs by students.

Thirdly, our exploration of the relationship between Perceived Trustworthiness, Procedural Justice, and Team Performance reveals problems in the construct of Team Trust in the current setting. The Team Trust measure utilized in our study, adapted from Costa and Anderson (2011), showed limitations, indicating the need for future refinement. The interplay between LLM usage, perceived Informational Justice, and Procedural Justice highlights the multifaceted nature of trust within collaborative learning environments. This finding suggests that as AI is entering the field of teamwork, theories might need to be adapted, maybe even from bidirectional to tridirectional relationships.

In light of these findings, the contribution of this study lies in its comprehensive examination of the intricate interplay between LLM Usage and lecturer-student trust dynamics. Although it needs to be regarded as preliminary first results and followed up with more statistical power, the identified nuances provide a foundation for future research and the development of guidelines for effective lecturer-student collaboration in the era of evolving technological, AI-powered tools. As universities worldwide grapple with integrating LLMs into educational practices, our study contributes valuable insights for shaping policies that foster ethical and transparent use, ensuring the continued trust and effectiveness of collaborative learning environments.

# Appendix

*Table 2:* Questionnaire Items

| Construct | Items | Scale | Adapted from |
|---|---|---|---|
| (Perceived) LLM Usage | LU1: I guess students use Large Language Models, such as ChatGPT, on a regular basis.<br>LU2: Large Language Models, such as ChatGPT, are not very common among students to work on assigned tasks. (R)<br>LU3: The use of Large Language Models, such as ChatGPT, among students is increasing rapidly, also within the context of their studies.<br>LU4: Students increasingly tend to use Large Language Models, such as ChatGPT, for their studies and research. | 5-point Likert scale<br><br>Strongly disagree<br>Rather disagree<br>Neutral<br>Rather agree<br>Strongly agree | Own scale |
| Informational Justice | IJ1: In my Lecturer-Student Collaboration, we provide each other with enough details for us to do our tasks.<br>IJ2: The information the student provides me about his work is sufficient and fair.<br>IJ3: In my Lecturer-Student Collaboration, we can trust the information that we share.<br>IJ4: In my Lecturer-Student Collaboration, students intentionally hide details on how they come up with solutions. (R) | Same as above | (Kernan & Hanges, 2002; Palanski et al., 2011) |
| Procedural Justice | PJ1: In my Lecturer-Student Collaboration, the methods students apply to come up with their solutions are fair.<br>PJ2: Students use unfair methods to come up with their solution. (R)<br>PJ3: In my Lecturer-Student Collaboration, I cannot trust the way a student came up with a solution. (R) | Same as above | (Colquitt, 2004; Dayan & Di Benedetto, 2010) |
| Propensity to Trust | PT1: Most students do not hesitate to help a person in need.<br>PT 2: The typical student is sincerely concerned about the problems of others.<br>PT3: Students usually tell the truth, even when they know they will be better off by lying. | Same as above | (Costa & Anderson, 2011) |
| Perceived Trustworthiness | PW1: In my Lecturer-Student Collaboration, we can rely on each other.<br>PW2: I have complete confidence in students' ability to perform tasks.<br>PW3: In my Lecturer-Student Collaboration, students may have hidden agendas. (R)<br>PW4: In my Lecturer-Student Collaboration, I guess students will keep their word. | Same as above | (Costa & Anderson, 2011) |
| Cooperative Behaviour | CB1: With students, I work in a climate of cooperation.<br>CB2: In my Lecturer-Student Collaboration, some students hold back relevant information in our collaboration. (R)<br>CB3: In my Lecturer-Student Collaboration, we discuss and deal with issues or problems openly.<br>CB4: In my Lecturer-Student Collaboration, students minimize what they tell about themselves. (R) | Same as above | (Costa & Anderson, 2011) |
| Monitoring Behaviour (R) | MB1: In my Lecturer-Student Collaboration, I need to watch students' work and methods very closely.<br>MB2: In my Lecturer-Student Collaboration, I check whether students keep their promises.<br>MB3: In my Lecturer-Student Collaboration, I should keep their work under surveillance. | Same as above | (Costa & Anderson, 2011) |
| Expected Team Performance | TP1: In my Lecturer-Student Collaboration, students put considerable effort into their jobs.<br>TP2: In my Lecturer-Student Collaboration, students are committed to producing quality work.<br>TP3: In my Lecturer-Student Collaboration, students meet or exceed their productivity requirements. | 5-point Likert scale<br><br>(The original scale was a 19-point scale. We adapted for randomization purposes.) | (Alper et al., 1998) |
| Manipulation Checks | MC1: When I read this question carefully, I select "neutral".<br>MC2: In case I answer all questions truthfully, I select "strongly agree". | See above | |
| Demographics and Controls | C1: What is your age? | 18–24, 25–34, 35–44, […], 75 years and older | |
| | C2: What is your gender? | Male, Female (question was not mandatory) | |
| | C3: How often do you use Large Language Models, such as ChatGPT? | Never, Barely, From time to time, Often, Daily | |
| | C4: How many students have you supervised so far in Final Year Projects, or similar settings? | 0, 1–20, 21–40, 41–60, 61–80, more than 81 | |
| | C5: Have you made bad experiences with students using Large Language Models, such as ChatGPT, so far? | Yes, more than once<br>Yes, once<br>Never | |
| | C6: How would you assess your experience in research? | 5-point Likert scale from Very Low to Very High (+ descriptions) | |
| | C7: Which field describes your professional expertise best? | ICT, Social Sciences, … | |
| | C8: Do you think Large Language Models, such as ChatGPT, are a fair tool in Higher Education? | Yes, No | |